\documentclass[fleqn,usenatbib]{mnras}

\usepackage{newtxtext,newtxmath}

\usepackage[T1]{fontenc}

\DeclareRobustCommand{\VAN}[3]{#2}
\let\VANthebibliography\thebibliography
\def\thebibliography{\DeclareRobustCommand{\VAN}[3]{##3}\VANthebibliography}

\usepackage{graphicx}	
\usepackage{amsmath}	
\usepackage{fontawesome}
\usepackage{multirow}
\usepackage{float}
\usepackage[normalem]{ulem}

\newcommand{\hi}{H\textsc{i}\ }
\newcommand{\hinospace}{\textrm{H\textsc{i}}}
\newcommand{\omc}{\Omega_{\mathrm{c}}h^2}
\newcommand{\omb}{\Omega_{\mathrm{b}}h^2}

\newcommand{\be}{\begin{equation}}
\newcommand{\ee}{\end{equation}}
\newcommand{\bea}{\begin{eqnarray}}
\newcommand{\eea}{\end{eqnarray}}

\title[\hi intensity mapping: perturbation theory predictions]{Interferometric \hi intensity mapping: perturbation theory predictions and foreground removal effects}

\author[]{
Alkistis Pourtsidou$^{1,2,3}$\thanks{E-mail: alkistis.pourtsidou@ed.ac.uk}\\
$^{1}$Institute for Astronomy, The University of Edinburgh, Royal Observatory, Edinburgh EH9 3HJ, UK\\
$^{2}$Higgs Centre for Theoretical Physics, School of Physics and Astronomy,
The University of Edinburgh, Edinburgh EH9 3FD, UK\\
$^{3}$ Department of Physics \& Astronomy, University of the Western Cape, Cape Town 7535, South Africa
}

\date{Accepted XXX. Received YYY; in original form ZZZ}

\pubyear{2022}

\begin{document}
\label{firstpage}
\pagerange{\pageref{firstpage}--\pageref{lastpage}}
\maketitle

\begin{abstract}
We provide perturbation theory predictions for the \hi intensity mapping power spectrum multipoles using the Effective Field Theory of Large Scale Structure (EFTofLSS), which should allow us to exploit mildly nonlinear scales. Assuming survey specifications typical of proposed interferometric \hi intensity mapping experiments like CHORD and PUMA, and realistic ranges of validity for the perturbation theory modelling, we run mock full shape MCMC analyses at $z=0.5$, and compare with Stage-IV optical galaxy surveys. We include the impact of 21cm foreground removal using simulations-based prescriptions, and quantify the effects on the precision and accuracy of the parameter estimation. We vary \textcolor{black}{11} parameters in total: 3 cosmological parameters, 7 bias and counterterms parameters, \textcolor{black}{and the \hi brightness temperature}. Amongst them, the 4 parameters of interest  are: the cold dark matter density, $\omega_{\rm c}$, the Hubble parameter, $h$, the primordial amplitude of the power spectrum, $A_{\rm s}$, and the linear \hi bias, $b_1$. For the best case scenario, we obtain unbiased constraints on all parameters with $<3\%$ errors at $68\%$ confidence level. \textcolor{black}{When we include the foreground removal effects, the parameter estimation becomes strongly biased for $\omega_{\rm c}, h$, and $b_1$, while $A_{\rm s}$ is less biased ($< 2\sigma$). We find that scale cuts $k_{\rm min} \geq 0.03 \, h/{\mathrm{Mpc}}$ are required to return accurate estimates for $\omega_{\rm c}$ and $h$, at the price of a decrease in the precision, while $b_1$ remains strongly biased. We comment on the implications of these results for real data analyses.}
\end{abstract}

\begin{keywords}
cosmology: theory -- cosmology: observations -- large-scale structure of the Universe -- methods:statistical
\end{keywords}



\section{Introduction}
\label{sec:Introduction}

Over the next few years, observations of the
redshifted 21cm line emission from neutral
hydrogen gas (\hinospace) with a new generation of radio telescopes can push the boundaries of our understanding of cosmology and galaxy evolution.
Remarkably, \hi
surveys can trace the matter distribution from the present time ($z=0$) to
the Epoch of Reionization ($z\sim 10$) and beyond, mapping a large part of the observable volume of the Universe. 

In the meantime, spectroscopic optical galaxy surveys have already proven extremely successful at mapping the low redshift Universe, and providing exquisite constraints on dark energy and gravity \citep[see, for example,][]{Mueller:2016kpu, Alam:2020sor}.   
These surveys operate by detecting galaxies in 3D, i.e., by measuring the redshift and angular position of each galaxy very precisely. 
In the radio wavelengths, due to the weakness of the \hi signal, being competitive with optical galaxy surveys using the traditional approach of detecting individual galaxies is extremely challenging. This challenge gave rise to an alternative observational technique, dubbed \hi intensity mapping, which maps the entire \hi flux coming from many galaxies together in 3D voxels. \citep{Battye:2004re,Chang:2007xk,Peterson:2009ka,Seo:2009fq,Ansari:2011bv}. Provided several observational challenges and systematic effects are mitigated or controlled, the \hi intensity mapping method has the potential to provide detailed maps of the Universe back to $\sim$1 billion years after the Big Bang \citep{Ahmed:2019ocj, Kovetz:2019uss, Moresco:2022phi}. 

A number of \hi intensity mapping experiments are expected to come online over the coming years, with some of them already taking data with pilot surveys. Examples are the proposed MeerKLASS survey using the South African MeerKAT array \citep{Santos:2017qgq}, FAST \citep{Hu:2019okh}, CHIME \citep{Bandura:2014gwa}, HIRAX \citep{Newburgh:2016mwi, Crichton:2021hlc}, Tianlai \citep{Li:2020ast, Wu:2020jwm}, PUMA \citep{PUMA:2019jwd}, and CHORD \citep{Vanderlinde:2019tjt}. Pathfinder surveys with the Green Bank Telescope (GBT), Parkes, CHIME, and MeerKAT, have achieved detections of the cosmological 21 cm emission, but have relied on cross-correlation analyses with optical galaxy surveys \citep{Chang:2010jp, Masui:2012zc, Anderson:2017ert, Wolz:2021ofa, CHIME:2022kvg, Cunnington:2022uzo}. 

A major challenge for the \hi intensity mapping method is the presence of strong astrophysical emission: 21cm foregrounds such as galactic synchrotron \citep{Zheng:2016lul}, point sources, and free-free emission, contaminate the maps and can be orders of magnitude stronger than the cosmological \hi signal \citep{Oh:2003jy}. Hence, they have to be removed. We can differentiate these dominant foregrounds
from the signal taking advantage of their spectral smoothness \citep{Liu:2011hh, Chapman:2012yj, Wolz:2013wna, Shaw:2014khi, Alonso:2014dhk, Cunnington:2020mnn}. As an example,
21cm foreground removal studies using low redshift \hi intensity mapping simulations and real data employ blind foreground removal techniques like Principal Component Analysis (PCA) \citep{Switzer_2013, Switzer_2015,Alonso:2014dhk} or Independent Component Analysis \citep{Hyvrinen1999FastAR, Wolz:2015lwa}. The procedure of foreground removal results in \hi signal loss, removing power from modes along the parallel and perpendicular to the line-of-sight (LoS) directions, with the parallel to the LoS effect  being more severe than the perpendicular one \citep{Witzemann:2018cdx, Cunnington:2020mnn}. 

The aim of this work is to investigate the systematic biases from 21cm foreground removal assuming state-of-the-art interferometric \hi intensity mapping experiments. To quantify how these systematic biases propagate on the cosmological parameter estimation, we will model the \hi signal using perturbation theory and run full shape MCMC analyses on synthetic data contaminated with 21cm foreground removal effects. We will also benchmark our predictions against a ``Stage-IV'' spectroscopic optical galaxy survey like DESI \citep{DESI:2016fyo} or Euclid \citep{EUCLID:2011zbd, Euclid:2019clj}.

The paper is organised as follows: In \autoref{sec:model} we present the perturbation theory model we will use. In \autoref{sec:mocks} we produce our synthetic mock data and contaminate them with simulated 21cm foreground removal effects. We present the results of our full shape MCMC runs in \autoref{sec:mcmc}. In \autoref{sec:conclusions} we summarise our findings and conclude.

\section{Theoretical modelling}
\label{sec:model}

Our observable is the \hi power spectrum multipoles, and we follow the formalism used in optical galaxy surveys analyses. Similarly to optical galaxies, redshift space distortions (RSDs) introduce anisotropies in the observed \hi power spectrum. In order to account for this, we consider the 3D power spectrum as a function of redshift $z$, $k$, and $\mu$, where $k$ is the amplitude of the wave vector and $\mu$ the cosine of the angle between the wave vector and the LoS component. This gives $k_\perp = \sqrt{1-\mu^2}$ and $k_\parallel = k\mu$.

Before we present the 1-loop perturbation theory model we will use, it is useful to discuss linear theory. We can model RSDs by considering the Kaiser effect \citep{Kaiser:1987qv}, which is a large-scale effect dependent on the growth rate, $f$. To linear order, the anisotropic \hi power spectrum can be written as: 
\begin{equation}
    \label{eq:linearmodel}
	P_\hinospace(k, \mu) =  \left( \overline{T}_\hinospace b_1 + \overline{T}_\hinospace f \mu^{2}\right)^{2} P_\text{m}(k) + P_\text{SN} + P_\text{N} \;\; .
\end{equation}
Here, $P_\text{m}(k)$ is the underlying matter power spectrum, $b_1$ is the (linear) \hi bias, and $\overline{T}_\hinospace$ is the mean \hi brightness temperature. $P_\text{N}$ is the  thermal noise of the telescope and $P_\text{SN}$ is the shot noise, $P_\text{SN} = \overline{T}_\hinospace^2 (1/\overline{n})$, where $\overline{n}$ is the number density of objects.
%
The $P_{\rm SN}$ contribution is expected to be subdominant (smaller than the thermal noise of the telescope) and is usually neglected \citep{Villaescusa-Navarro:2018vsg}. The noise power spectrum for a typical interferometer is given by \citep{Zaldarriaga:2003du, Bull:2014rha}:
\begin{equation}
\label{eq:Pnoise_IM}
    P_{\mathrm{N}}=T_{\mathrm{sys}}^{2} r^{2} y_{\nu}\left(\frac{\lambda^{4}}{A_{\mathrm{e}}^{2}}\right) \frac{1}{2 n(u=k_\perp r/2\pi) t_{\mathrm{total}}}\left(\frac{S_{\mathrm{area}}}{\mathrm{FOV}}\right) \;\; .
\end{equation}
Here, $A_e$ is the effective beam area, ${\rm FOV} \approx \lambda / (D_{\rm dish})^2$, $r$ is the comoving distance to the observation redshift $z$, and $y_\nu = c(1+z)^2/(\nu_0H(z))$ with $\nu_0 = 1420$ MHz. $T_{\mathrm{sys}}$ is the system temperature, $S_{\mathrm{area}}$ is the survey area, and $t_{\mathrm{total}}$ is the total observing time.
 
\textcolor{black}{The antennae distribution function $n(u)$ can be calculated using a fitting formula \citep{CosmicVisions21cm:2018rfq}. For a square array with $N^2_{\rm s}$ receivers, the number of baselines as a function of physical distance of antennas is given by
\begin{equation}
    n^{\rm phys}_{\rm b}(l) = n_0 \frac{a + b(l/L)}{1+c(l/L)^d}
    e^{-(l/L)^e} \, ,
\end{equation} 
where $n_0 = (N_{\rm s}/D_{\rm dish})^2$, $L=N_{\rm s}D_{\rm dish}$, and the $uv$-plane density is 
\begin{equation}
\label{eq:nu_dist}
    n(u) = \lambda^2 n^{\rm phys}_{\rm b}(l=u\lambda) \, . 
\end{equation} 
}

The \hi abundance and clustering properties have been studied using simulations and semi-analytical modeling \citep[see, e.g.,][]{Padmanabhan:2016fgy,Villaescusa-Navarro:2018vsg,Spinelli:2019smg}. 
The clustering of \hi should be accurately described by perturbative methods at mildly nonlinear scales \citep{10.1093/mnras/stw1111, McQuinn:2018zwa, Sarkar:2019nak, Castorina:2019zho, Sailer:2021yzm,  Karagiannis:2022ylq, Qin:2022xho}. Modelling nonlinear scales is necessary in order to get precise and accurate cosmological constraints with instruments like HIRAX, CHIME, CHORD, and PUMA, and it also helps break degeneracies, for example between $b_1$ and the primordial power spectrum amplitude, $A_{\mathrm{s}}$. Similar degeneracies exist for $\overline{T}_\hinospace$, which is proportional to the \hi mean density, $\Omega_{\hinospace}$. Accurate ($<5\%$) measurements of $\Omega_{\hinospace}(z)$ are available at low redshifts \citep{Crighton:2015pza}, and it can also be constrained by joint analyses of different probes \citep{Obuljen:2017jiy,Chen:2018qiu} or by exploiting very small scales that can be described using bespoke \hi halo models \citep{Chen:2020uld}.

In this work we will use the ``EFTofLSS'' formalism that has been developed to model the power spectrum multipoles of biased tracers in redshift space.
The main difference between this model and the standard 1-loop Standard Perturbation Theory (SPT) formalism \citep{Bernardeau:2001qr} is that the EFTofLSS approach accounts for the impact of nonlinearities on mildly nonlinear scales by introducing effective
stresses in the equations of motion. This results in the addition of counterterms to the 1-loop
power spectrum, which represent the effects of short distance
physics at long distances.

The EFTofLSS model we will employ is described in various papers (see e.g. \citet{Perko:2016puo} and references therein), and we refer to \citet{DAmico:2019fhj} for its application to the DR12 BOSS data. Main assumptions are that we live in a spatially expanding, homogeneous and isotropic background spacetime, and that we work on sub-horizon scales with $\delta, \theta \ll 1$ ( where $\delta$ and $\theta$ are the density and velocity
perturbations, respectively). 

The 1-loop redshift space galaxy power spectrum then reads \citep{Perko:2016puo, DAmico:2019fhj}:

\begin{align}
P_{g}(k, \mu) &=Z_{1}(\mu)^{2} P_{11}(k) \nonumber \\
&+2 \int \frac{d^{3} q}{(2 \pi)^{3}} Z_{2}(\boldsymbol{q}, \boldsymbol{k}-\boldsymbol{q}, \mu)^{2} P_{11}(|\boldsymbol{k}-\boldsymbol{q}|) P_{11}(q) \nonumber \\
&+6 Z_{1}(\mu) P_{11}(k) \int \frac{d^{3} q}{(2 \pi)^{3}} Z_{3}(\boldsymbol{q},-\boldsymbol{q}, \boldsymbol{k}, \mu) P_{11}(q) \nonumber \\
&+2 Z_{1}(\mu) P_{11}(k)\left(c_{\mathrm{ct}} \frac{k^{2}}{k_{\mathrm{M}}^{2}}+c_{r, 1} \mu^{2} \frac{k^{2}}{k_{\mathrm{M}}^{2}}+c_{r, 2} \mu^{4} \frac{k^{2}}{k_{\mathrm{M}}^{2}}\right) \nonumber \\
&+\frac{1}{\bar{n}_{g}}\left(c_{\epsilon, 1}+c_{\epsilon, 2}   \frac{k^{2}}{k_{\mathrm{M}}^{2}}+c_{\epsilon, 3} f \mu^{2} \frac{k^{2}}{k_{\mathrm{M}}^{2}}\right) \, , \label{eq:Pgkmu}
\end{align}
where $k_{\rm M} = 0.7h$/Mpc and $\bar{n}_{\rm g}$ is the mean galaxy density. The various terms are summarised nicely in \citet{Nishimichi:2020tvu}, and we follow this description here: the $c_{\rm ct}$ term represents a  combination of a higher derivative bias and the speed of
sound of dark matter; the $c_{r,i}$ terms represent the redshift-space counterterms, while the $c_{\epsilon,i}$ terms represent the stochastic counterterms. The kernels $Z_1$, $Z_2$, and $Z_3$ are the redshift-space galaxy density kernels appearing in the 1-loop power spectra. They are expressed in terms of the galaxy density and velocity kernels and 4 bias parameters: \{$b_1,b_2,b_3,b_4$\}. 
For flat $\Lambda$CDM, which we will assume in this work, the logarithmic growth rate $f$ is calculated by solving for the linear growth factor $D$ (with $a$ the scale factor), and yields: 
$
f(a)=\frac{(5 a-3 D(a)) \Omega_{m}}{2 D(a)\left(\Omega_{m}+a^{3}\left(1-\Omega_{m}\right)\right)}.
$

The model of \autoref{eq:Pgkmu} has recently been implemented in a publicly available \texttt{Python} code, \texttt{PyBird} \citep{DAmico:2020kxu}. 
In principle, the model can describe any biased tracer of matter, so we can straightforwardly apply it to \hinospace. Following the literature we perform the following changes of variables: $b_2 = \frac{1}{\sqrt{2}}(c_2+c_4)$, $b_4 = \frac{1}{\sqrt{2}}(c_2-c_4)$, $c_{\epsilon, \rm mono} = c_{\epsilon,1} + \frac{1}{3}c_{\epsilon,2}$, $c_{\epsilon, \rm quad} = 
    \frac{2}{3} c_{\epsilon,2}$,
and also fix $c_4=c_{r,2}=c_{\epsilon,\rm mono}=0$ so that our final set of nuisance parameters is: $\{ b_1, c_2,b_3,c_{\rm ct},c_{\rm{r},1},c_{\epsilon,1},c_{\epsilon, \rm quad} \}$. We will comment on these choices when we construct our mock data in \autoref{sec:mocks}.

The cosmological parameters that the code takes as input are: the cold dark matter density $\omega_{\rm c} = \omc$, the baryonic matter density $\omega_{\rm b} = \omb$, the Hubble parameter $h$, the amplitude of the primordial
power spectrum, $A_{\rm s}$, and the scalar spectral index, $n_{\rm s}$. We will describe the code and other software we used to speed-up the parameter inference in more detail in \autoref{sec:mcmc}.

\section{Mock data}
\label{sec:mocks}

For our analysis we produce synthetic \hi monopole and quadrupole data running \texttt{Pybird} for a central redshift $z=0.5$. We will not use the hexadecapole as it is not expected to add significant cosmological information, and it is more affected by nonlinear uncertainties. In addition, as shown in \citet{Cunnington:2020mnn, Soares:2020zaq}, the \hi intensity mapping hexadecapole (as well as higher order multipoles) can be used for identifying the effects of foreground removal and other systematics. \textcolor{black}{Not using the hexadecapole allows us to set $c_{{\rm r}, 2} = 0$. The choice $c_4=c_{\epsilon,\rm mono}=0$ is motivated by the assumption that the functions multiplying $c_4$ and $c_{\epsilon,\rm mono}$ are too small to affect the results. These assumptions follow the BOSS data analyses choices \citep{DAmico:2019fhj}, but they will need to be reaffirmed with bespoke \hi simulations and real data.} 
The fiducial cosmological parameters are \citep{Aghanim:2018eyx}: 
$$
    \{ \omega_{\rm c}, h, A_{\rm s}, \omega_{\rm b}, n_{\rm s} \} =
    \\ \{ 0.1193, 0.677, 3.047, 0.0224, 0.967\} \, .
$$
\textcolor{black}{For setting the fiducial values of the nuisance parameters, we perform fits to \hi intensity mapping simulations. These are described in \autoref{sec:appendix-MD}, and we find:
$$
    \{ b_1, c_2,b_3,c_{\rm ct},c_{\rm{r},1},c_{\epsilon,1},c_{\epsilon, \rm quad} \} = 
    \\ \{ 1.1, 0.6 , 0.1, 0.1, -10,  0, -0.8 \} \, .
$$
We remark that the value of the linear \hi bias $b_1$ is in very good agreement with values found at similar redshifts in other works \citep{10.1093/mnras/stw1111, Villaescusa-Navarro:2018vsg}. We also note that in all our MCMC forecasts we will marginalise over the nuisance parameters.}

The model in \autoref{eq:Pgkmu} has to be rescaled by the square of the \hi brightness temperature, $\bar{T}_{\hinospace}(z)$, which in turn depends on the \hi abundance, $\Omega_{\hinospace}(z)$. Using the fitting function from \citet{SKA:2018ckk} we set $\bar{T}_{\hinospace}(z=0.5) = 0.168 \, {\rm mK}$ \textcolor{black}{as our fiducial value for this parameter}.

We also need a data covariance. To calculate this we will assume an ambitious CHORD-like intensity mapping survey \citep{Vanderlinde:2019tjt}. CHORD (Canadian Hydrogen Observatory and Radio transient Detector) is a successor to CHIME \citep{CHIME:2022dwe}, aiming to perform a very large sky \hi intensity mapping survey. Its core array consists of 512 dishes, each 6m in diameter. The bandwidth is large, covering the 300-1500 MHz band, or redshifts up to $z=6$. We will assume $T_{\rm sys}=50$ K in our forecasts. Another very ambitious proposal is PUMA, a close-packed interferometer array with 32,000 dishes, covering the
frequency range 200-1100 MHz, or redshifts $0.3 < z < 6$ \citep{PUMA:2019jwd}. We expect both of these instruments to be able to achieve similar signal-to-noise ratios ($S/N$), and we will focus on CHORD from now on. 

Aiming to establish how CHORD can complement and compete with state-of-the-art optical galaxy surveys, we choose a low redshift bin centred at $z=0.5$ with width $\Delta z = 0.3$.  Our fiducial CHORD-like survey covers $20,000 \, {\rm deg}^2$ on the sky, resulting in a survey bin volume $V_{\rm sur} = 4246^3 \, ({\rm Mpc}/h)^3$. We assume a $20,000$ hrs survey \textcolor{black}{and calculate the noise power spectrum using \autoref{eq:Pnoise_IM}, with the fitting parameters needed in \autoref{eq:nu_dist} being $a=0.4847, b=-0.3300, c=1.3157, d=1.5974, e=6.8390$ \citep{CosmicVisions21cm:2018rfq}. The corresponding baseline density $n(u)$ for our CHORD-like array is shown in \autoref{fig:nu_dist} (see Appendix A of \citet{Karagiannis:2022ylq} for the case of a HIRAX-like array).}
\begin{figure}
\centering
  \includegraphics[scale=0.55]{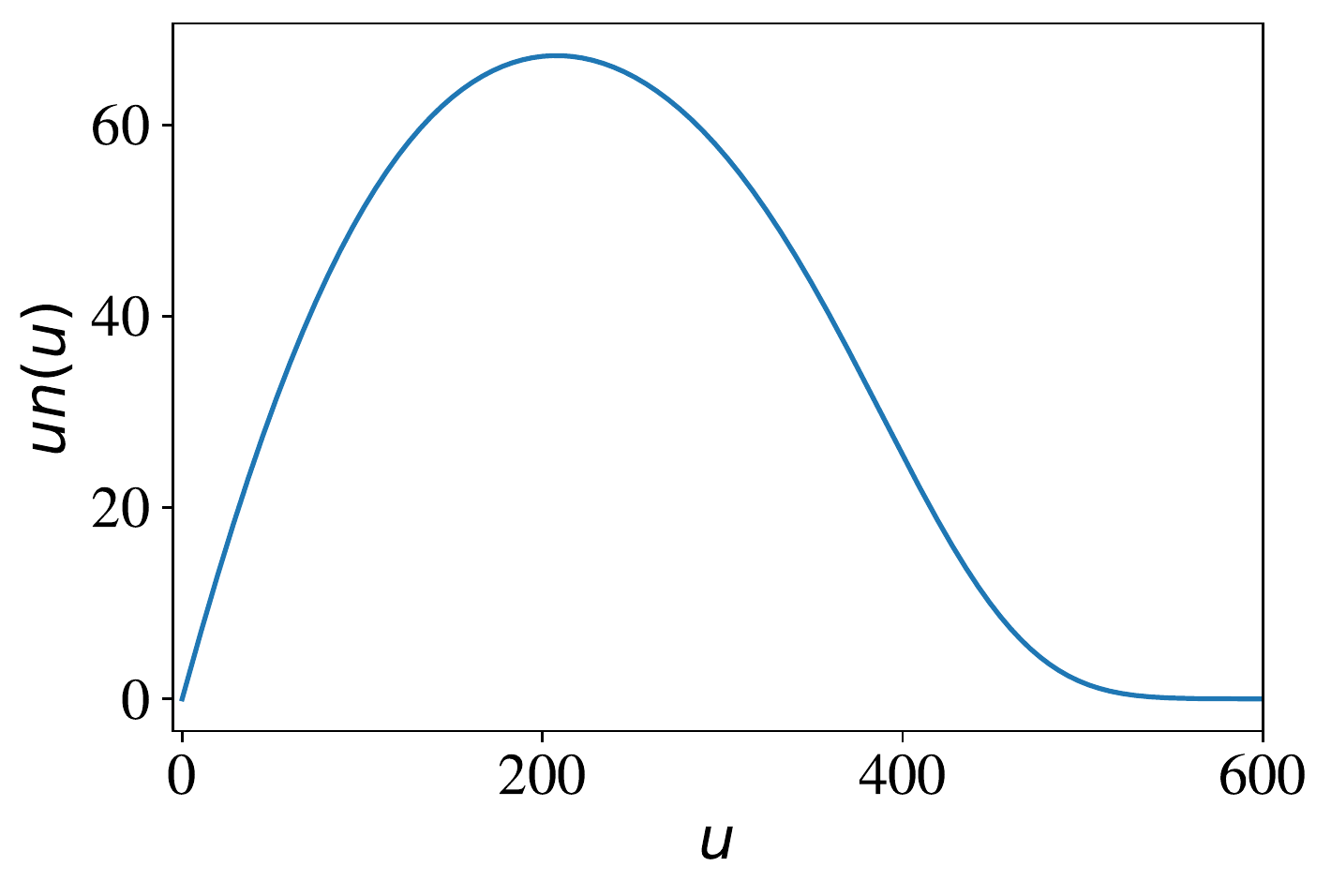}
  \caption{\textcolor{black}{Baseline density for our CHORD-like array, calculated using \autoref{eq:nu_dist}.}}
\label{fig:nu_dist}
\end{figure}

\begin{figure}
\centering
  \includegraphics[scale=0.55]{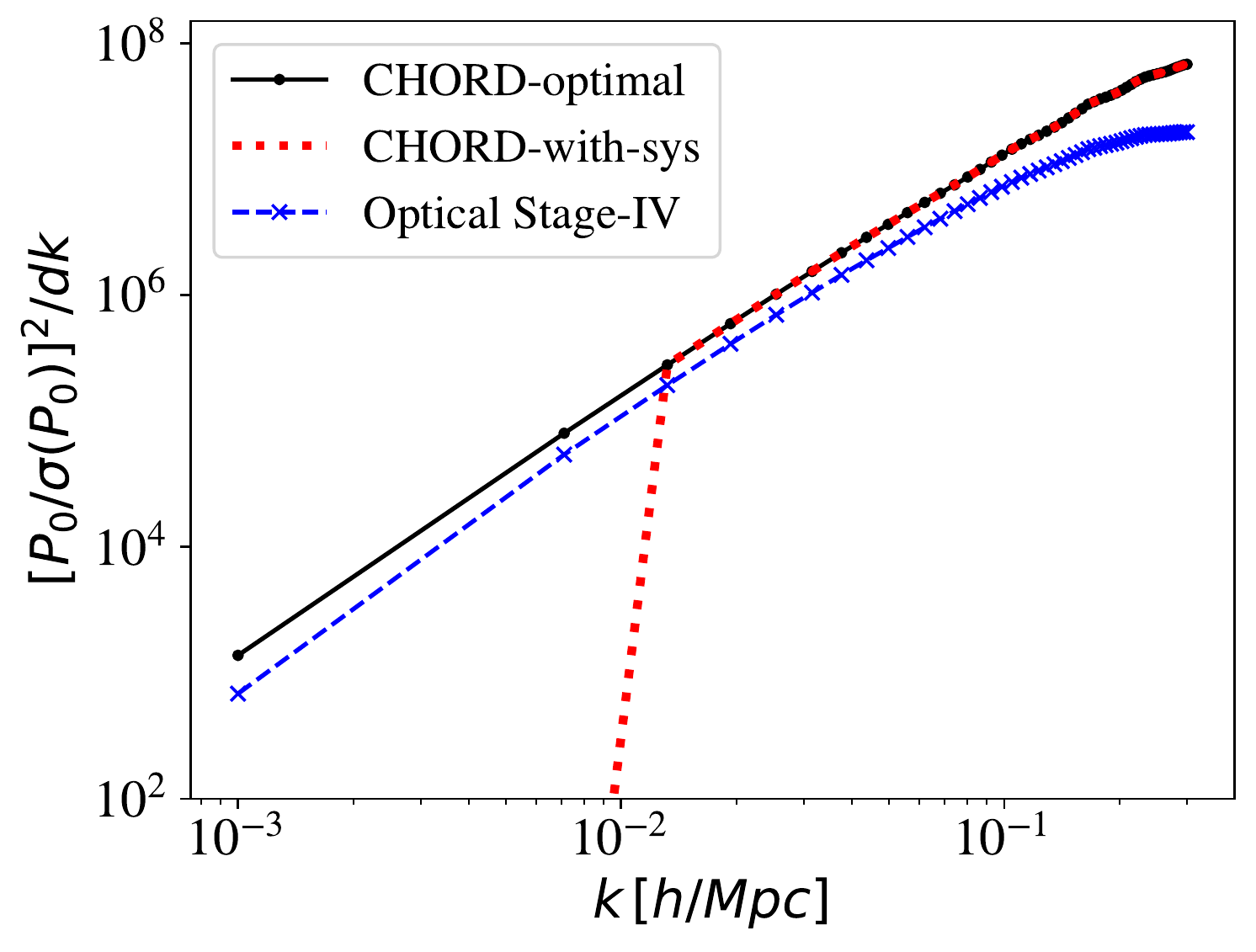}
  \caption{\textcolor{black}{Signal-to-noise-ratio (squared) for the spherically averaged power spectrum, $P_0$, for different surveys. ``Optical Stage-IV'' represents a spectroscopic galaxy survey with $\bar{n}_{\rm g} = 0.0005 \, h^3/{\rm Mpc}^3$. ``CHORD-optimal'' corresponds to an idealised \hi intensity mapping survey, while ``CHORD-with-sys'' illustrates the case where sensitivity is lost at small $k$.}}
\label{fig:SNR}
\end{figure}

\textcolor{black}{We can now calculate $P_{\mathrm{N}}$ for our CHORD-like survey using \autoref{eq:Pnoise_IM}. Dividing by $\bar{T}^2_{\rm HI}$ and inverting, we can define an effective mean number density $\bar{n}_{\rm HI}$. For a Stage-IV spectroscopic galaxy survey like DESI \citep{DESI:2016fyo} or Euclid \citep{EUCLID:2011zbd, Euclid:2019clj}, the shot noise is the inverse of the number density of galaxies, $\bar{n}_{\rm g}$.
In \autoref{fig:SNR} we plot the signal-to-noise-ratio (squared) for the power spherically averaged power spectrum (the monopole, $P_0$) for different surveys. Stage-IV corresponds to a spectroscopic optical galaxy survey with $\bar{n}_{\rm g} = 0.0005 \, h^3/{\rm Mpc}^3$. ``CHORD-optimal'' corresponds to an idealised case for an interferometric \hi intensity mapping survey without any systematic effects, while ``CHORD-with-sys'' illustrates the case where sensitivity is lost at small $k$ (see also Fig. 15 in \citep{CosmicVisions21cm:2018rfq}). This can be due to foreground removal effects which mainly affect the small $k_\parallel$, and/or inability to probe the small $k_\perp$ due to baseline restrictions. For the case of the CHORD-like survey at $z=0.5$, this can result to loss of sensitivity in a range $\sim 0.001 \leq k \leq 0.06 \, h/\mathrm{Mpc}$, and we will consider different $k_{\rm min}$ scale cuts in our forecasts to take this into account. In all cases, \autoref{fig:SNR} demonstrates that a CHORD-like experiment can achieve a higher signal-to-noise-ratio in the nonlinear regime, compared to a Stage-IV optical galaxy survey.}

\begin{figure}
\centering
  \includegraphics[scale=0.55]{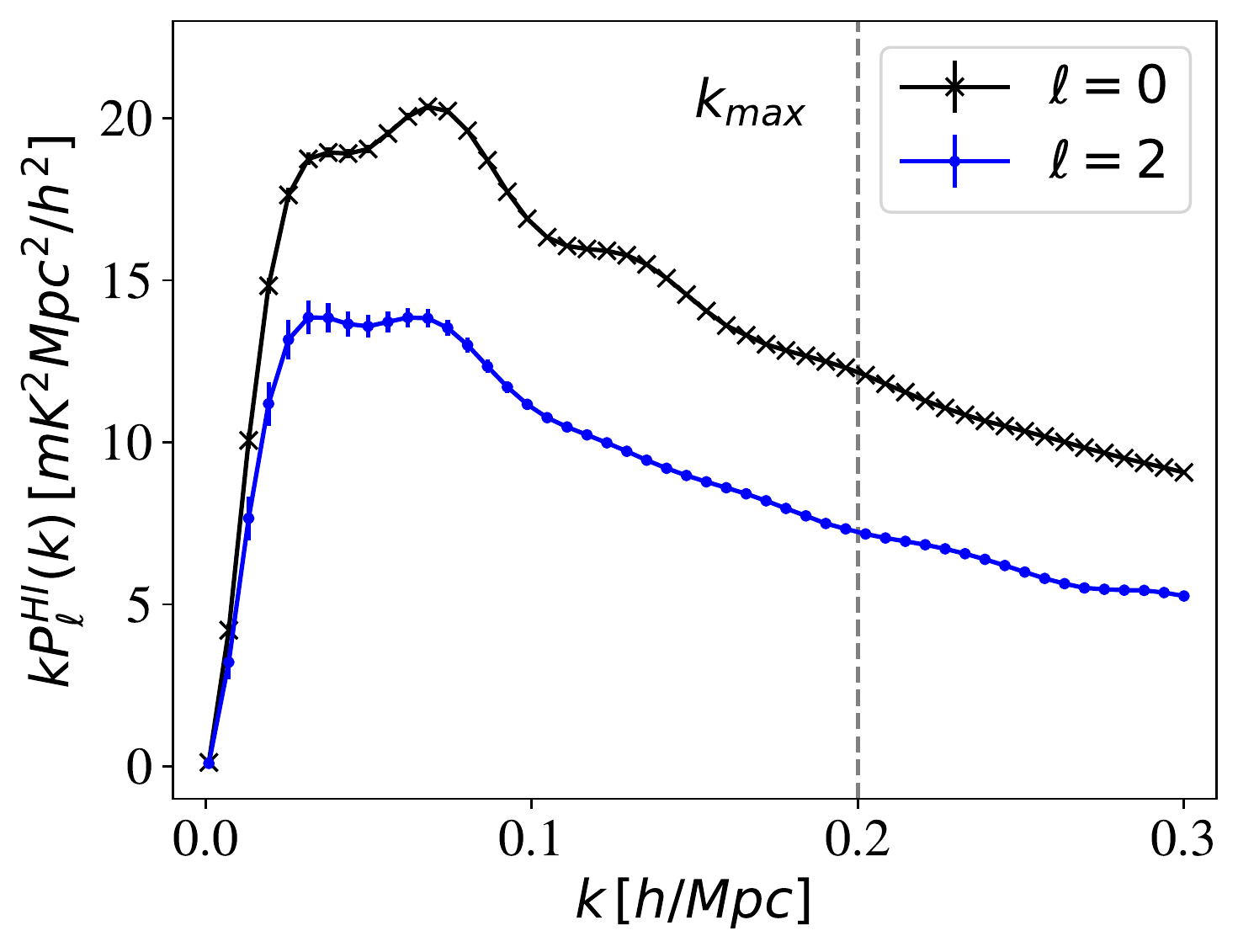}
  \caption{Our fiducial monopole ($\ell=0$) and quadrupole ($\ell=2$) data, assuming the \textcolor{black}{CHORD-optimal} \hi intensity mapping survey. The vertical dashed grey line denotes the maximum wavenumber (smallest scale used in the MCMC) $k_{\rm max} = 0.2h/$Mpc at our chosen central redshift $z=0.5$.}
\label{fig:mock-data}
\end{figure}

We can now proceed to calculate the multipole covariances analytically, using the Gaussian approximation \citep{Taruya:2010mx, Soares:2020zaq}. We present the resulting mock data and measurement errors in \autoref{fig:mock-data} for the \textcolor{black}{CHORD-optimal case}. We notice that in the case of the monopole the error bars are not large enough to be visible.
At this low redshift, and with such high $S/N$, nonlinear uncertainties are expected to become important at a relatively low $k$. Hence, we choose the range of validity for the EFTofLSS modelling to be $k < k_{\rm max} = 0.2\, h/$Mpc (this should be a good assumption, but it has to be validated with tailored \hi simulations for these experiments). We are now ready to perform MCMC forecasts.

\section{MCMC analyses}
\label{sec:mcmc}

To calculate posterior distributions on the parameters we have run MCMCs using the ensemble slice sampling codes \texttt{emcee} \citep{Foreman-Mackey:2012any} and \texttt{zeus} \citep{Karamanis:2021tsx}. The latter has been recently used to run mock full shape MCMC analyses assuming galaxy surveys specifications using the \texttt{Matryoshka}  suite of neural network based emulators \citep{Donald-McCann:2021nxc, Donald-McCann:2022pac}. Due to the impressive increase in computational speed for the inference ($\sim 3$ orders of magnitude improvement with respect to the \texttt{PyBird} runs), we opted for this setup to run and present our final MCMC forecasts\footnote{An alternative approach to speed-up the inference is to use a fast and accurate linear matter power spectrum emulator such as \texttt{bacco} \citep{Arico:2021izc} or \texttt{CosmoPower} \citep{SpurioMancini:2021ppk} as input in a perturbation theory code, instead of running a Boltzmann solver like \texttt{CAMB} \citep{Lewis:1999bs} or \texttt{CLASS} \citep{Lesgourgues:2011re, Blas:2011rf}. 
}. 
We vary three cosmological parameters, $\left[\omega_{\rm c}, h, \ln \left(10^{10} A_{\rm s}\right)\right]$, seven bias and counterterms parameters, $\left[b_1, c_2,b_3,c_{\rm ct},c_{\rm{r},1},c_{\epsilon,1},c_{\epsilon, \rm quad}\right]$, \textcolor{black}{and, in the case of IM, we also vary $\bar{T}_{\hinospace}$.} The scalar spectral index $n_{\rm s}$ is fixed to its true value, and so is the baryon fraction $f_{\rm b} = \omega_{\rm b} / (\omega_{\rm b} + \omega_{\rm c})$. 

\begin{figure*}
\centering
  \includegraphics[scale=0.4]{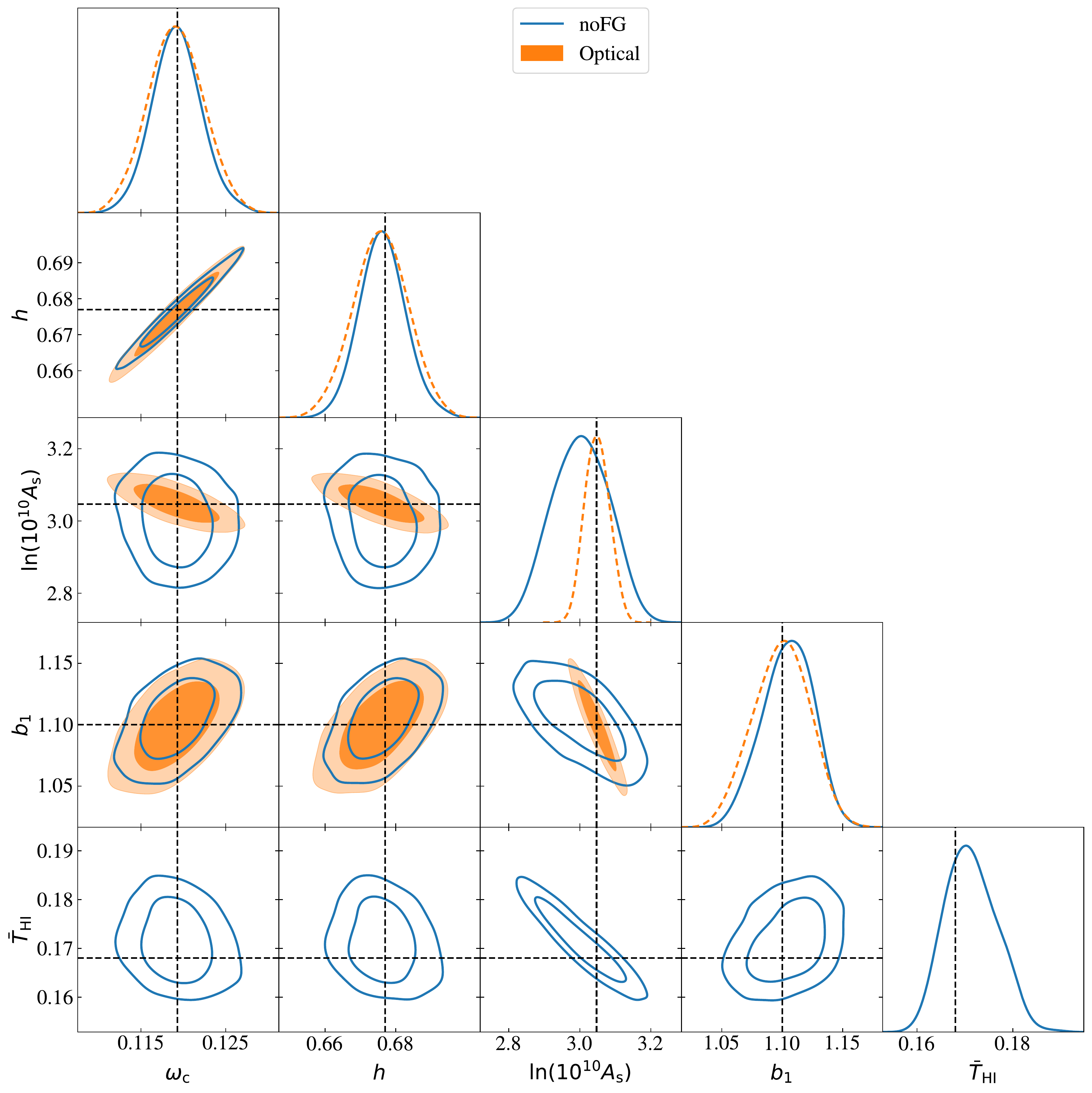}\\
  \caption{Marginalised 1D and 2D posterior distributions, and $1\sigma$ and $2\sigma$ contours, for a Stage-IV spectroscopic optical galaxy survey and a CHORD-like intensity mapping (IM) survey. The black dashed line shows the fiducial (true) parameters. \textcolor{black}{We show the 4 parameters of interest and $\bar{T}_{\hinospace}$, but we remind the reader that a total of 10
parameters are included in the MCMC for the optical survey, and 11 for the IM survey (see \autoref{sec:appendix}, \autoref{fig:MCMCfull}).}}
\label{fig:Stage-IV-opt-vs-IM}
\end{figure*}

For the 3 cosmological parameters and $b_1$, we assume the uniform flat priors shown in \autoref{tab:constraints}. We do not employ Planck priors on $A_{\rm s}, \omega_{\rm c},h$ because we wish to assess the precision vs accuracy performance of interferometric \hi intensity mapping independently of CMB experiments. 
\textcolor{black}{For $\bar{T}_{\hinospace}$ we take a flat prior $[0,1]$.} 
For the rest of the bias and counterterms parameters, we follow \citet{DAmico:2019fhj}
and set:
\begin{align*} 
&c_2\sim\mathcal{U}(-4, 4),
&b_3\sim\mathcal{N}(0, 2), \\ &c_{ct}\sim\mathcal{N}(0, 2),
&c_{r,1}\sim\mathcal{N}(0, 8), \\ &c_{\epsilon,1}\ / \ n_{\hinospace} \sim\mathcal{N}(0, 400),
&c_{\rm quad}\sim\mathcal{N}(0, 2).
\end{align*}

Finally, we assume a Gaussian likelihood given by: 
\be
\ln \mathcal{L}(P^{\rm d} \mid \theta)=-\frac{1}{2}(P^{\rm d}-{P}^{\rm m})^{T} \boldsymbol{C}^{-1}(P^{\rm d}-{P}^{\rm m}) \, ,
\ee
with $P^{\rm d}$ being the mock data (the power spectrum monopole, $P_0$, and quadrupole, $P_2$), $P^{\rm m}$ being the EFTofLSS model predictions
for a given set of parameters, $\theta$, and $\boldsymbol{C}$  being the covariance matrix.

\subsection{The systematics-free, Stage-IV survey scenario}

We start by comparing the performance of a Stage-IV spectroscopic galaxy survey and an analogous intensity mapping survey, assuming both of them are free of systematic effects. The volumes of the surveys are taken to be exactly the same\footnote{\hi intensity mapping surveys can cover a much wider redshift range compared to spectroscopic optical galaxy surveys, but our goal here is to compare their performance on a given redshift bin.}, \textcolor{black}{but the effective mean number densities (i.e., the noise components) are different as we have described in detail in \autoref{sec:mocks} (see e.g. \autoref{fig:SNR}).}
Following up on the discussion in the previous section, we emphasise again that in the case of \hi intensity mapping there is an additional overall amplitude parameter, $\bar{T}^2_{\hinospace} \propto \Omega^2_{\hinospace}$, \textcolor{black}{which we vary}.  
\textcolor{black}{This means that a total of 10 (11) parameters are varied in the MCMC for the optical (IM) surveys under consideration.}

The MCMC contours for the idealised case are shown in \autoref{fig:Stage-IV-opt-vs-IM}. We are able to recover the true values in an unbiased manner, which also confirms the accuracy of the \texttt{Matryoshka} emulator (see \citet{Donald-McCann:2022pac} for a suite of validation tests). 
The Stage-IV spectroscopic optical galaxy survey and the CHORD-like \hi intensity mapping survey have similar constraining power when the same survey volume is assumed. \textcolor{black}{An exception is the primordial amplitude $A_{\rm s}$. In the CHORD-like IM case, additional degeneracies due to varying $\bar{T}_{\hinospace}$ increase the uncertainty in $A_{\rm s}$ compared to the optical case.} 

From \autoref{tab:constraints} we see that, in the absence of systematics, the CHORD-like \hi intensity mapping survey determines $\omega_{\mathrm{c}}$ with $<3\%$ error, $h$ with $1\%$ error, \textcolor{black}{$\ln \left(10^{10} A_{\rm s}\right)$ with $<3\%$ error}, and $b_1$ (the linear \hi bias) with $<2\%$ error. \textcolor{black}{The survey can also constrain $\bar{T}_{\hinospace} = 0.171^{+0.005}_{-0.006}$, demonstrating how exploiting mildly nonlinear scales can break degeneracies}.

\subsection{Contaminating the data vector with 21cm foreground removal effects}

In order to contaminate our synthetic data vector (i.e., the \hi power spectrum multipoles $P_0$ and $P_2$ in \autoref{fig:mock-data}) with 21cm foreground removal effects, we will use the simulations-based prescription by \citet{Soares:2020zaq}. This prescription can fit \hi intensity mapping simulations with foreground removal effects, assuming that PCA or FastICA with $N_{\rm IC}=4$ independent components  was used for the foreground cleaning. The choice $N_{\rm IC}=4$ corresponds to an excellent calibration scenario (which will hopefully be the case by the time CHORD and PUMA come online) and no polarization leakage \citep{Wolz:2013wna, Alonso:2014dhk, Liu:2019awk, Cunnington:2020njn}. For existing \hi intensity mapping surveys, we know that much more aggressive foreground removal (much higher $N_{\rm IC} \sim 20-30$) is employed to deal with more complicated foregrounds, noise, and unknown systematics (see e.g. \citet{Switzer_2013,Masui:2012zc, Wolz:2021ofa, Cunnington:2022uzo}).

\begin{figure}
\centering
  \includegraphics[scale=0.4]{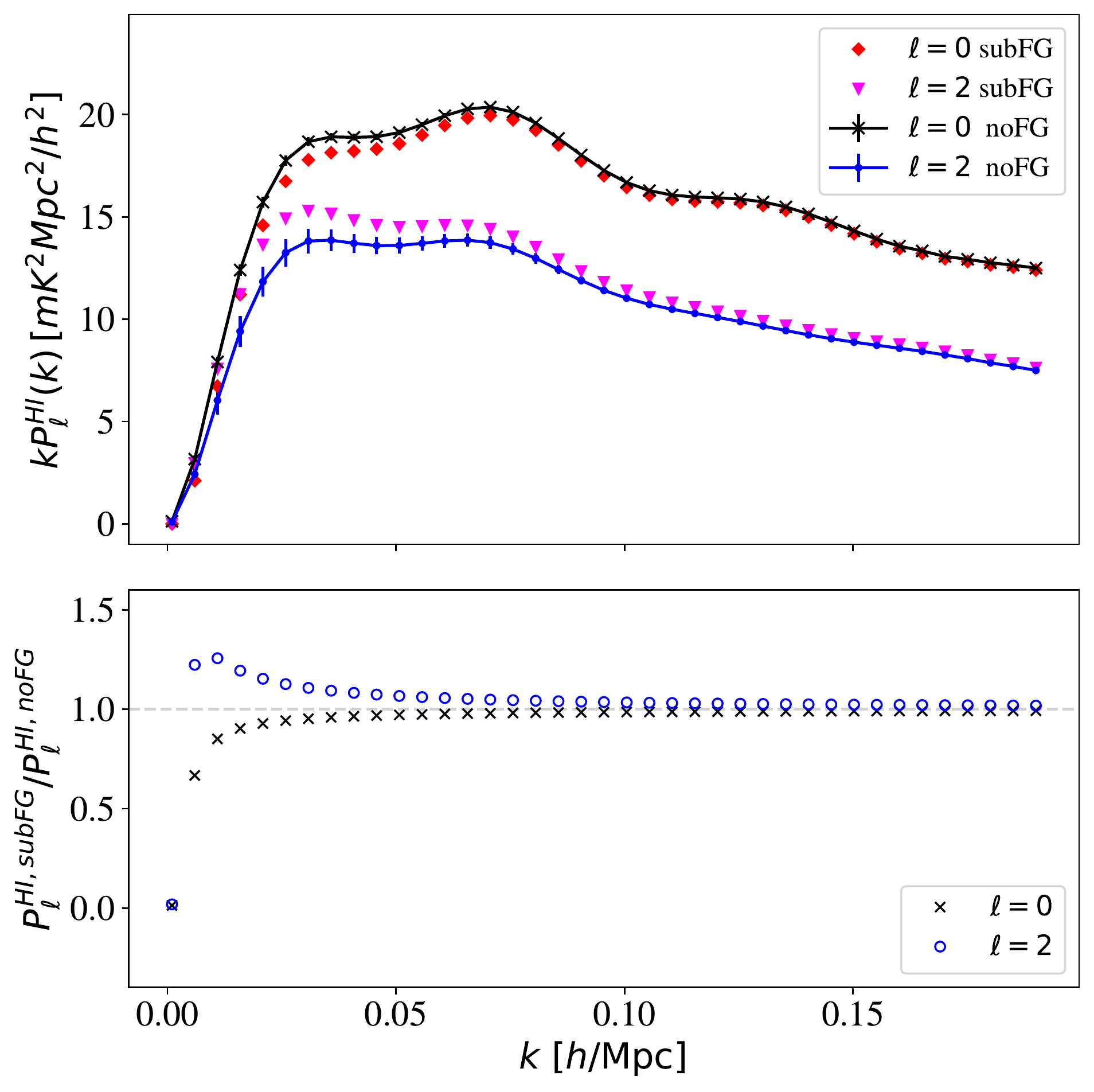}
  \caption{\textit{Top}: Fiducial monopole ($\ell=0$) and quadrupole ($\ell=2$) data, with and without foreground removal effects, assuming a CHORD-like intensity mapping survey. \textit{Bottom}: The ratios of the power spectrum multipoles illustrate the amplitude and scale dependence of the foreground removal effects.}
\label{fig:mock-data-with-FG}
\end{figure}

We present the result of contaminating our data vector with 21cm foreground removal effects in \autoref{fig:mock-data-with-FG}. This is our mock data vector for the remainder of the paper.  
As we can see, foreground cleaning results in the damping of power across a wide range of scales in the spherically averaged power spectrum (the monopole, $P_0$). This is a well known effect, which has also been identified in the context of high redshift 21cm surveys of the Epoch of Reionization \citep{Petrovic_2011}. Higher order multipoles were first studied extensively in \citet{Blake:2019ddd, Cunnington:2020mnn, Soares:2020zaq}, focusing on post-reionization \hinospace. In the case of the quadrupole ($P_2$), where $P(k,\mu)$ is weighted as a function of $\mu$, we see an enhancement of power on large scales. It is also important to note that, both for $P_0$ and $P_2$, while the largest effect is clearly on small $k$, there is still an effect along larger $k$. Given that the error bars of our chosen surveys are extremely small, the large $k$ effect may introduce a significant systematic bias as well. We can only verify and quantify this at the level of the parameter inference, and we will do so in the following sections.

We remark that we assume no prior knowledge of the 21cm foreground removal fitting function from \citet{Soares:2020zaq}. That is, we will not attempt to include a model (and associated nuisance parameters we need to vary) for the 21cm foreground removal effects in our theory vector. 
We will instead follow a more conservative, ``data-driven'' approach, imposing scale cuts on the contaminated data vector. However, the former method has been shown to be very promising in the context of single-dish experiments, and it would be important to re-evaluate its performance when realistic, end-to-end simulations are available (see e.g. the discussion in \citet{Spinelli:2021emp}). 

\begin{figure}
\centering
  \includegraphics[width=0.45\textwidth]{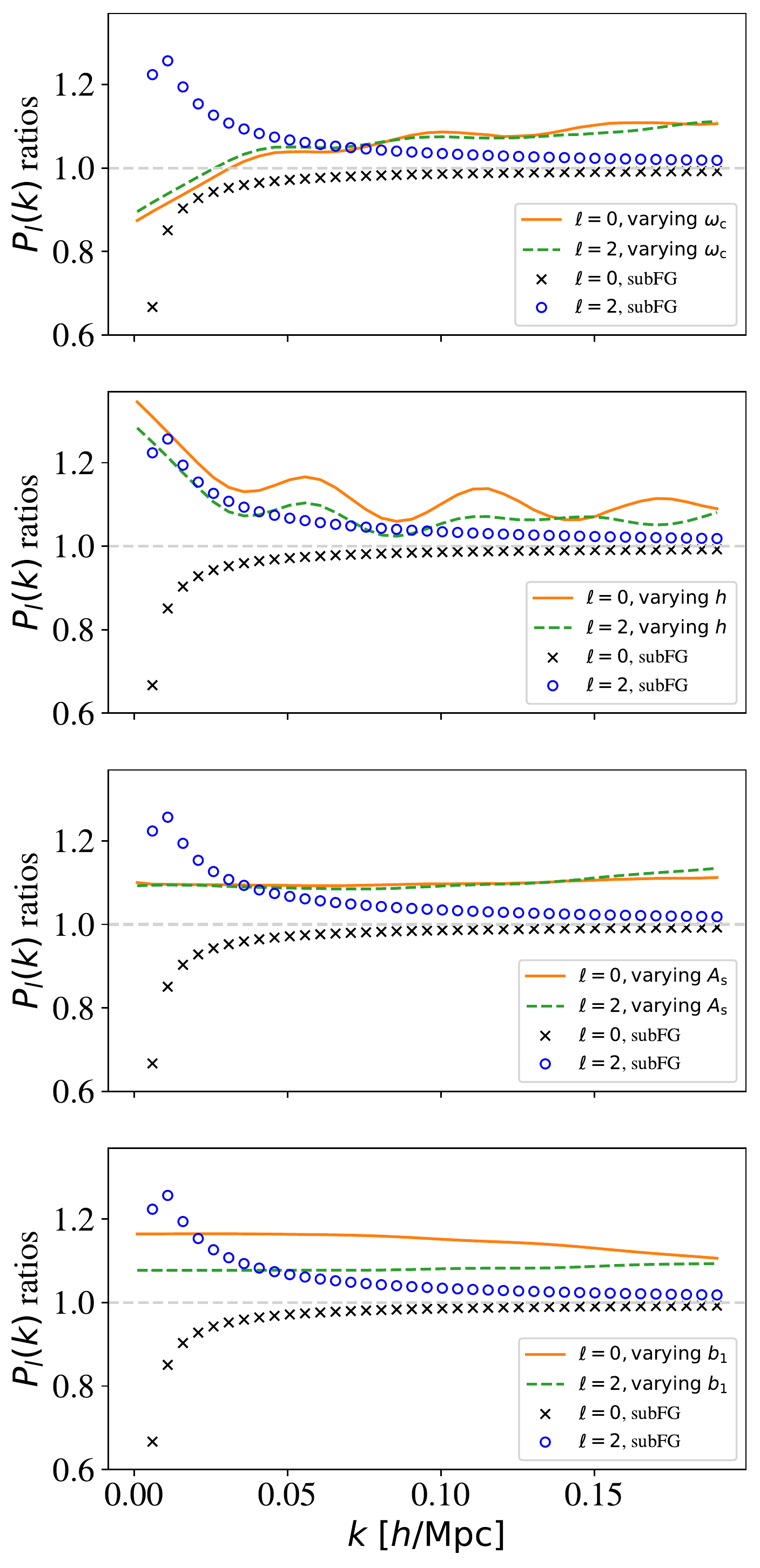}
  \caption{The effect of $\sim 10\%$ variations of the 4 parameters of interest on the EFTofLSS model predictions. For comparison, we also plot the simulations-based effect of 21cm foreground removal.}
\label{fig:vary-params}
\end{figure}

Finally, before presenting our MCMC analyses using the contaminated data vector, it is instructive to see how varying our 4 parameters of interest affects the predictions of the EFTofLSS model of \autoref{eq:Pgkmu}, and compare with the features of the 21cm foreground removal effects. 
This comparison is shown in \autoref{fig:vary-params}. 
It is well known that different cosmological and nuisance parameters affect the power spectrum amplitude and shape in distinct ways. Comparing these with our simulated 21cm foreground removal effects suggests that some parameters, for example $h$, have features on the mildly nonlinear scales that might make them easier to constrain in an unbiased way (i.e., to disentangle them from the foreground removal effects) than others, for example $A_{\mathrm{s}}$ and $b_1$.

\subsection{Imposing $k_{\mathrm{min}}$ cuts: precision vs accuracy}

\begin{figure}
\centering
  \includegraphics[width=0.47\textwidth]{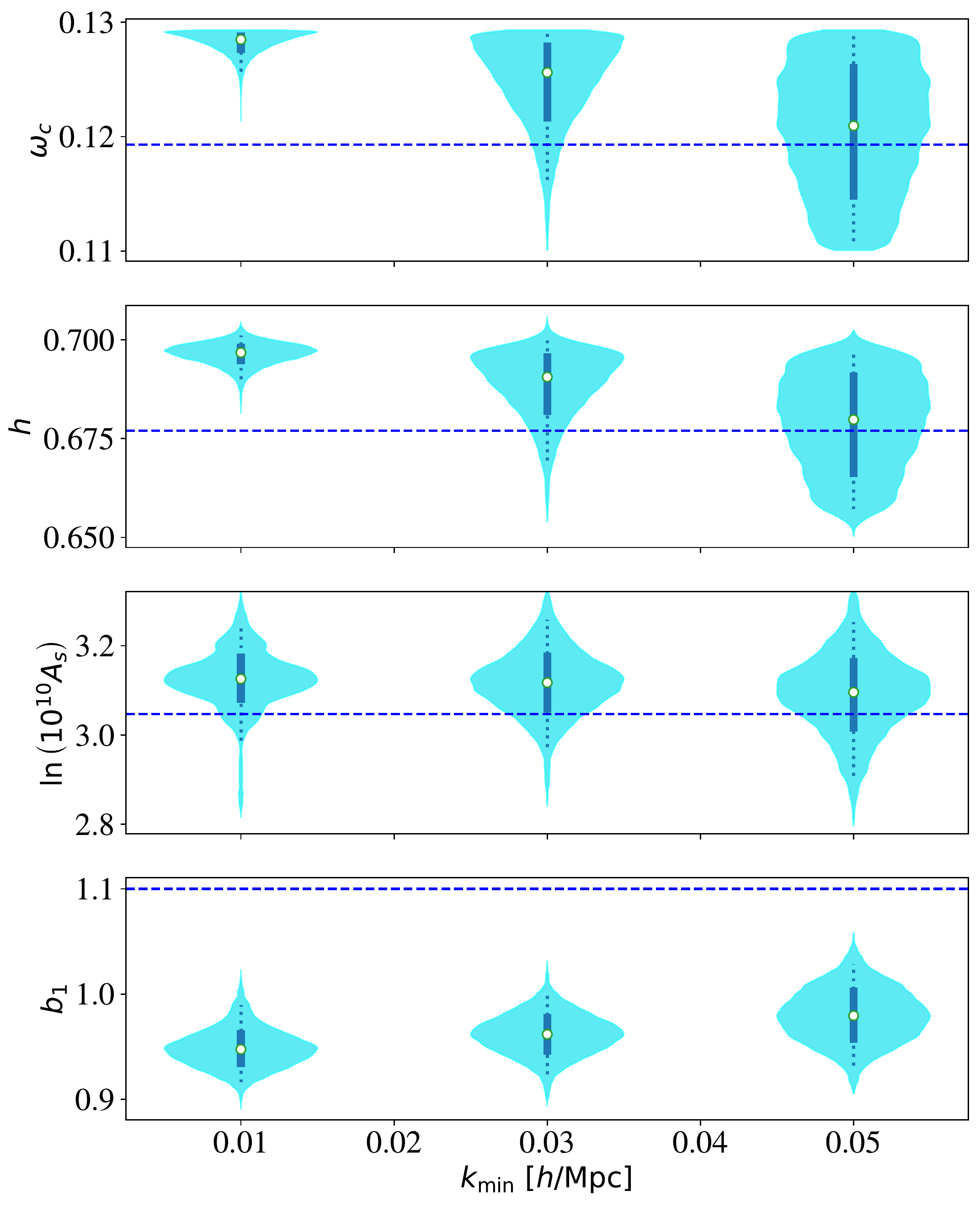}
  \caption{Violin plots showing the marginalised 1D posteriors on the 4 parameters of interest for the different $k_{\rm min}$ limits we have considered for the CHORD-like IM survey \textcolor{black}{contaminated by foreground removal effects}. The dashed
blue lines show the true (fiducial) values. The white points show the median values. The thick solid (thin dotted) blue lines show the $1\sigma$ ($2\sigma$) regions. The shaded cyan regions show the density of the marginalised
posteriors. We remind the reader that a total of 11
parameters are included in the MCMC.}
\label{fig:violins}
\end{figure}

In this section we perform an MCMC analysis for the CHORD-like survey under consideration, using the contaminated \hi intensity mapping data vector with different $k_{\rm min}$ limits. Our results for the 4 parameters of interest are summarised in \autoref{tab:constraints} and \autoref{fig:violins}. The different scale cuts we consider are the following:
\begin{itemize}

    \item \textit{Case I}: We start by imposing a scale cut $k_{\rm min} = 0.01 \, h/{\mathrm{Mpc}}$ in order to exclude the largest scales where foreground subtraction has the most impact. This is not enough: \textcolor{black}{the parameter estimation becomes strongly biased for $\omega_{\rm c}, h$, and $b_1$. The primordial amplitude is unbiased within $< 2\sigma$ because of the marginalisation of  $\bar{T}_{\hinospace}$ (i.e., if $\bar{T}_{\hinospace}$ was kept fixed, $A_{\rm s}$ would also be strongly biased). For the \hi brightness temperature we get: $\bar{T}_{\hinospace} = 0.173^{+0.004}_{-0.004}$.}
    
    \item \textit{Case II}: Imposing a stricter scale cut $k_{\rm min} = 0.03 \, h/{\mathrm{Mpc}}$. \textcolor{black}{In this case the $\omega_{\rm c}$, $h$, and $A_{\rm s}$ parameters are unbiased within $2\sigma$, while $b_1$ remains biased. For the \hi brightness temperature we get: $\bar{T}_{\hinospace} = 0.174^{+0.004}_{-0.005}$.}
    
    \item \textit{Case III}: Imposing a scale cut $k_{\rm min} = 0.05 \, h/{\mathrm{Mpc}}$. \textcolor{black}{In this case the $\omega_{\rm c}$, $h$, and $A_{\rm s}$ parameters are unbiased within $1\sigma$, while $b_1$ remains biased. For the \hi brightness temperature we get: $\bar{T}_{\hinospace} = 0.176^{+0.004}_{-0.006}$.}
    
\end{itemize}

\begin{figure}
\centering
  \includegraphics[scale=0.4]{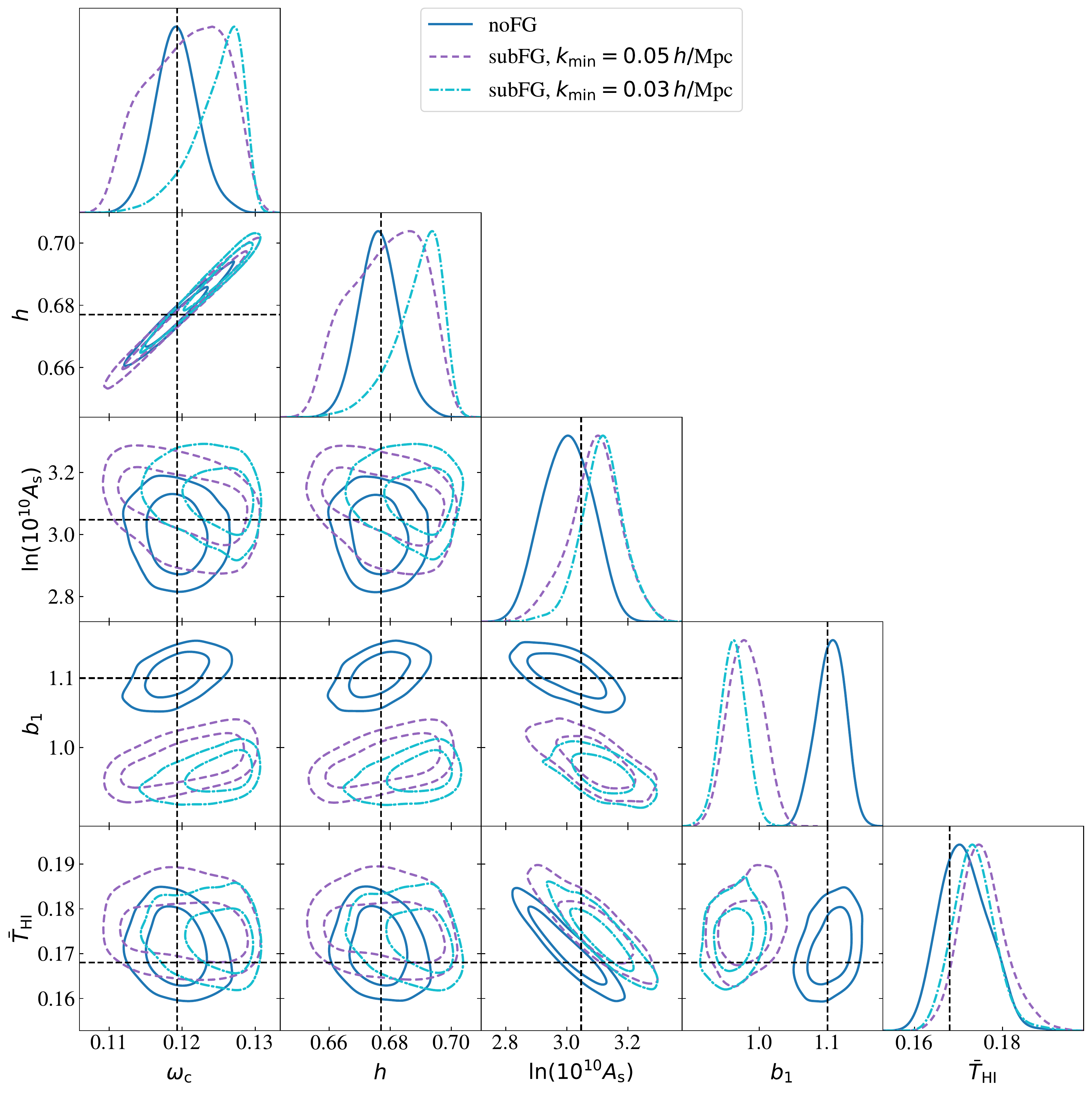}\\
  \caption{Marginalised 1D and 2D posterior distributions, and $1\sigma$ and $2\sigma$ contours for a CHORD-like \hi intensity mapping survey. We compare the idealised case and the cases with \textcolor{black}{ $k_{\rm min} \geq 0.03 \, h/\mathrm{Mpc}$ limits that lead to unbiased (within $2\sigma$) constraints on $\omega_{\rm c}$, $h$, and $A_{\rm s}$}. The black dashed line shows the fiducial (true) parameters. We show the 4 parameters of interest and \textcolor{black}{$\bar{T}_{\hinospace}$}, but we remind the reader that a total of \textcolor{black}{11}
parameters are included in the MCMC (see \autoref{sec:appendix}, \autoref{fig:MCMCfull}).}
\label{fig:subFG_kmin}
\end{figure}

In \autoref{fig:subFG_kmin} we compare the idealised case (noFG) and the cases with \textcolor{black}{ $k_{\rm min} \geq 0.03 \, h/\mathrm{Mpc}$ limits that lead to unbiased (within $2\sigma$) constraints on $\omega_{\rm c}$, $h$, and $A_{\rm s}$}. Looking back at \autoref{fig:vary-params}, we deduce that for $k_{\rm min} \geq 0.03 \, h/{\mathrm{Mpc}}$ the scale-dependent features of varying $\omega_{\rm c}$ and $h$ are sufficient to constrain them accurately, in contrast to $b_1$, which gets significantly biased due to the amplitude change from the 21cm foreground removal. \textcolor{black}{The primordial amplitude $A_{\rm s}$ is less affected because of the marginalisation of  $\bar{T}_{\hinospace}$ (we have checked that when $\bar{T}_{\hinospace}$ is kept fixed, $A_{\rm s}$ becomes strongly biased).}

From \autoref{tab:constraints}, \autoref{fig:violins} and \autoref{fig:subFG_kmin} we also see that the price to pay for the unbiased estimates of $\omega_{\rm c}$ and $h$ parameters is a decrease in the precision with respect to the idealised case, as expected.

\section{Conclusions}
\label{sec:conclusions}

We have provided perturbation theory predictions for \hi intensity mapping, and performed full shape MCMC analyses including the impact of 21cm foreground removal.
Albeit our framework was developed in the context of low-redshift, interferometric \hi intensity mapping surveys like CHIME, HIRAX, CHORD, and PUMA, our results are also relevant for single-dish surveys as well as Epoch of Reionization surveys. The main conclusions we draw from this work are:

\begin{itemize}

    \item In the idealised case without any systematic biases in the data, interferometric \hi intensity mapping surveys with instruments like CHORD and PUMA are competitive with Stage-IV optical galaxy surveys. Our results, summarised in \autoref{tab:constraints}, assume a single redshift bin centred at $z=0.5$. In the case of redshift-independent parameters like $\omega_{\rm c}, A_{\rm s}$, and $h$, we \textcolor{black}{naively} expect the parameter estimation uncertainties to be reduced by roughly $1/\sqrt{N}$, where $N$ is the number of redshift bins. This is an advantage for \hi intensity mapping surveys with respect to optical surveys, since the former are not shot-noise limited and can rapidly cover very large redshift ranges. \textcolor{black}{However, the thermal noise of an interferometer can increase rapidly with redshift. Nevertheless, forecasts for experiments similar to what we consider here have shown that interferometric \hi intensity mapping should be advantageous at high redshift $2<z<6$. The cosmological volume spanned at this range is three times higher than the typical optical galaxy surveys range, $0<z<2$, with an increased $k_{\rm max}$, i.e., an increased number of (easier to model) linear and mildly nonlinear modes. Forecasts for a dedicated ``Stage-II'' \hi intensity mapping experiment showed that $S/N >1$ is achievable for all modes with $k \leq 0.4 h/{\rm Mpc}$ at $z \leq 6$ \citep{CosmicVisions21cm:2018rfq}.} 
    
    \item Including 21cm foreground removal effects based on simulations-based prescriptions, \textcolor{black}{the parameter estimation becomes strongly biased for $\omega_{\rm c}, h$, and $b_1$, while $A_{\rm s}$ is less biased ($< 2\sigma$). }
    
    \item Adopting the scale cuts approach to try and mitigate the biases, \textcolor{black}{we find that scale cuts $k_{\rm min} \geq 0.03 \, h/{\mathrm{Mpc}}$ are required to return accurate estimates for $\omega_{\rm c}$ and $h$, at the price of a decrease in the precision, while $b_1$ remains biased.}

\end{itemize}

In future work it would be interesting to investigate the possible $\hinospace$ and cosmology dependence of the foreground transfer function. This is a method to correct for signal loss from foreground removal, which has been used in all the \hinospace-galaxy cross-correlation detections so far.  Due to the low $S/N$ of current experiments, the cosmology is kept fixed to the Planck best-fit values and the only parameter we can constrain is $\Omega_{\hinospace}b_{1}r$, with $r$ the \hinospace-galaxy cross-correlation coefficient \citep{Masui:2012zc, Anderson:2017ert, Wolz:2021ofa, Cunnington:2022uzo}. The foreground transfer function is constructed using mock simulations with fixed \hi and cosmological parameters. In light of our results, we believe it is important to study how robust the transfer function construction (and the resulting \hi signal loss correction) is with respect to varying the parameters in the mock simulations. 

\onecolumn

\begin{table}
\centering
\begin{tabular}{||c c c c c c ||} 
\hline
 Parameters of interest & & $\omega_{\mathrm{c}}$ & $h$ & $\ln \left(10^{10} A_{\rm s}\right)$ & $b_1$ \\
 \hline
 Fiducial values & & 0.1193 & 0.677 & 3.047 & \textcolor{black}{1.1} \\
 Priors & & [0.101,0.140] & [0.575,0.748] & [2.78,3.32] & [0,4] \\
 \hline\hline
 \hline
 Case & $k$-range [$h/$Mpc] &  &  &  &  \\ 
 \hline
 \textcolor{black}{Optical galaxy survey} (\autoref{fig:Stage-IV-opt-vs-IM}) & $0.001 < k < 0.2$  & $0.119^{+0.003}_{-0.003}$ & $0.676^{+0.008}_{-0.008}$ & $3.05^{+ 0.04}_{-0.04}$ & $1.1^{+ 0.03}_{-0.02}$ \\
 \hline
 \textcolor{black}{IM-noFG} (\autoref{fig:Stage-IV-opt-vs-IM}) & $0.001 < k < 0.2$  & $0.119^{+0.003}_{-0.003}$ & $0.676^{+0.007}_{-0.007}$ & $3.00^{+ 0.08}_{-0.08}$ & $1.1^{+ 0.02}_{-0.02}$ \\ 
 \hline
 \textcolor{black}{IM-subFG, $k_{\rm min}= 0.01 \, h/\mathrm{Mpc}$} (\autoref{fig:violins}) & $0.01 < k < 0.2$  & $0.1282^{+0.0011}_{-0.0005}$ & $0.697^{+0.003}_{-0.002}$ & $3.12^{+ 0.06}_{-0.05}$ & $0.95^{+ 0.02}_{-0.02}$ \\
 \hline
 \textcolor{black}{IM-subFG, $k_{\rm min}= 0.03 \, h/\mathrm{Mpc}$} (\autoref{fig:violins} and \autoref{fig:subFG_kmin}) & $0.03 < k < 0.2$  & $0.125^{+0.004}_{-0.002}$ & $0.689^{+0.010}_{-0.005}$ & $3.12^{+0.07}_{-0.07}$ & $0.96^{+ 0.02}_{-0.02}$ \\
 \hline
 \textcolor{black}{IM-subFG, $k_{\rm min}= 0.05 \, h/\mathrm{Mpc}$} (\autoref{fig:violins} and \autoref{fig:subFG_kmin}) & $0.05 < k < 0.2$  & $0.121^{+0.007}_{-0.004}$ & $0.679^{+0.015}_{-0.010}$ & $3.09^{+ 0.09}_{-0.07}$ & $0.98^{+ 0.03}_{-0.03}$ \\
 \hline
\end{tabular}
\caption{Fiducial values, prior ranges, and marginalised constraints for the 4 parameters of interest (68\% confidence level). \textcolor{black}{We vary 10 parameters in total for optical galaxy surveys, and 11 for intensity mapping (IM) (adding the $\bar{T}_{\hinospace}$ parameter). For the IM surveys, we consider different $k_{\rm min}$ limits to mitigate 21cm foreground removal effects.}}
\label{tab:constraints}
\end{table}

\twocolumn

\section*{Acknowledgements}
I acknowledge use of open source software:
\texttt{Python} \citep{van1995python,Hunter:2007},  \texttt{numpy} \citep{numpy:2011},
\texttt{scipy} \citep{2020SciPy-NMeth},
\texttt{TensorFlow}  \citep{tensorflow2015-whitepaper}, 
\texttt{astropy} \citep{astropy:2018}, \texttt{corner} \citep{corner}, 
\texttt{GetDist} \citep{Lewis:2019xzd}. I thank New Mexico State
University (USA) and Instituto de Astrofisica de Andalucia CSIC
(Spain) for hosting the Skies \& Universes site for cosmological
simulation products. 
My research is funded by a UK Research and Innovation Future Leaders Fellowship [grant MR/S016066/1]. For the purpose of open access, the author has applied a Creative Commons Attribution (CC BY) licence to any Author Accepted Manuscript version arising from this submission.

\section*{Data Availability}

The data products will be shared on reasonable request to the corresponding author. 



\bibliographystyle{mnras}
\bibliography{main} 




\appendix

\section{Fits to simulations}
\label{sec:appendix-MD}

\textcolor{black}{Here we fit the EFTofLSS model of \autoref{eq:Pgkmu} to \hi simulations in order to determine fiducial values for our bias and counter terms parameters. The simulations we use have been described in detail in \citet{Soares:2021ohm}, but we summarise them here for completeness. They are based on the \texttt{MULTIDARK-PLANCK} dark matter N-body simulation \citep{Klypin:2014kpa}, which follows $3840^3$ particles evolving in a box of side $1 \, {\rm Gpc}/h$. The cosmology is consistent with \texttt{PLANCK15} \citep{Planck:2015fie}. From this, the \texttt{MULTIDARK-SAGE} catalogue was produced by applying the semi-analytical model \texttt{SAGE} \citep{Knebe:2017eei,Croton:2016etl}. These simulated data products are publicly available in the Skies \& Universe\footnote{\url{http://skiesanduniverses.org/}} web page, and we use the $z=0.39$ snapshot. The method used for generating \hi intensity mapping simulations is as follows: Each galaxy in the \texttt{MULTIDARK-SAGE} catalogue has an associated cold gas mass, which is used to calculate an \hi mass. The \hi mass of each galaxy belonging to each voxel is binned together, and then converted into a \hi brightness temperature for that voxel. A crucial limitation comes from the fact that low mass halos (lower than $\leq 10^{10} h^{-1}M_{\odot}$) are not included in the simulation. To account for these missing halos, which would contribute to the total \hi brightness of each voxel, we need to rescale the mean \hi temperature of the simulation to a more realistic value. In the power spectrum measurements, this is an overall amplitude $\bar{T}^2_{\hinospace}$, which we have divided out for the purposes of this fit.}

\begin{figure}
\centering
  \includegraphics[scale=0.5]{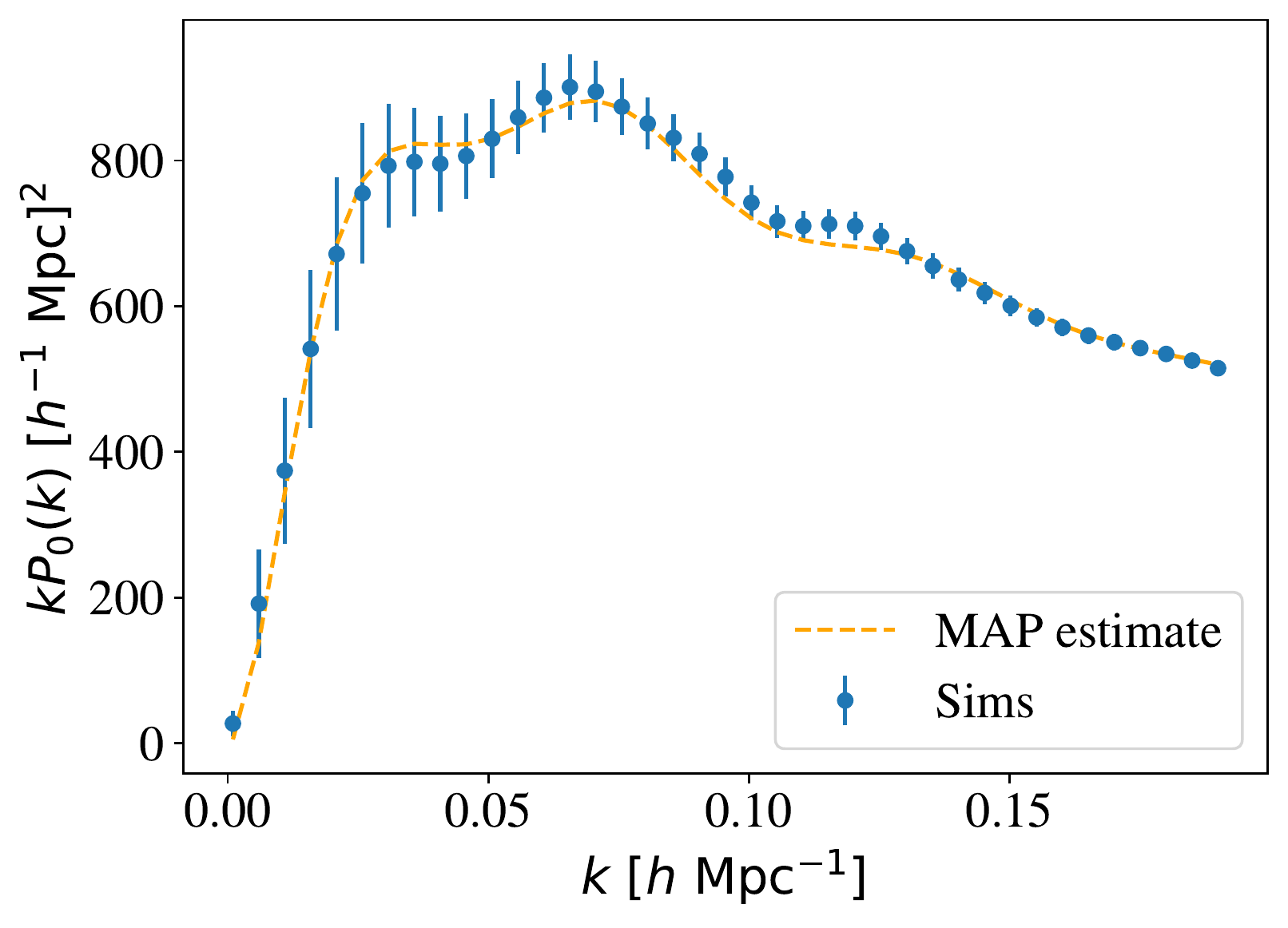}\\
  \caption{Comparison of the power spectrum monopole ($P_0$) predictions at the maximum a posteriori (MAP) estimate to the simulations measurements.}
\label{fig:MultiDark-fit}
\end{figure}

\textcolor{black}{We then proceed to find the maximum a posteriori (MAP) estimate to the power spectrum measurements from these simulations. We plot the result in \autoref{fig:MultiDark-fit}. The MAP values of the bias and counter terms parameters are: 
$$
    \{ b_1, c_2,b_3,c_{\rm ct},c_{\rm{r},1},c_{\epsilon,1},c_{\epsilon, \rm quad} \} = 
    \\ \{ 1.1, 0.6 , 0.1, 0.1, -10,  0, -0.8 \} \, .
$$
}

\section{Full posteriors}
\label{sec:appendix}

\onecolumn

\begin{figure}
\centering
  \includegraphics[scale=0.3]{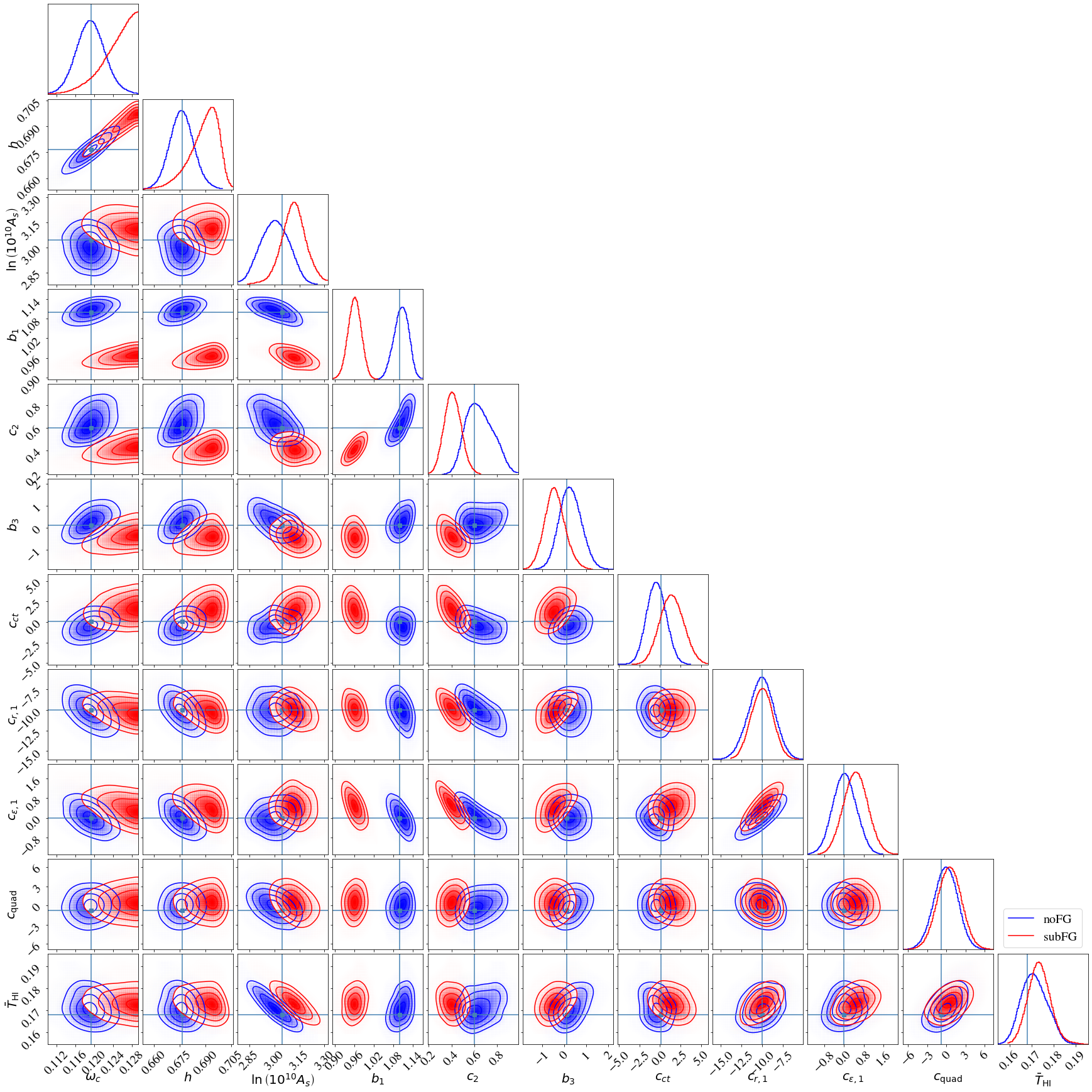}\\
  \caption{Marginalised 1D and 2D posterior distributions, and $1\sigma$ and $2\sigma$ contours, for a CHORD-like \hi intensity mapping survey, including all \textcolor{black}{11} parameters varied in the MCMC. We show the idealised case (noFG) and a case with 21cm foreground removal effects and a scale cut $k_{\rm min} = 0.03 \, h/{\mathrm{Mpc}}$ (subFG).}
\label{fig:MCMCfull}
\end{figure}


\bsp	
\label{lastpage}
\end{document}